\documentclass[a4paper,fleqn,usenatbib]{mnras}

\usepackage{newtxtext,newtxmath}

\usepackage[T1]{fontenc}
\usepackage{ae,aecompl}
\usepackage{etoolbox}
\makeatletter
\patchcmd\@combinedblfloats{\box\@outputbox}{\unvbox\@outputbox}{}{%
   \errmessage{\noexpand\@combinedblfloats could not be patched}%
}%
 \makeatother

\usepackage{graphicx}	
\usepackage{amsmath}	
\usepackage{amssymb}	

\newcommand {\kms} {\,{\rm km\,s}^{-1}}

\newcommand {\mo}{{\rm M}_\odot}

\newcommand{\aref}[1]{\hyperref[#1]{Appendix~\ref{#1}}}

\title[Mixing of metals during star cluster formation]{Mixing of metals during star cluster formation: statistics and implications for chemical tagging}

\author[L. Armillotta, M. R. Krumholz and Y. Fujimoto]{Lucia Armillotta$^{1}$\thanks{E-mail: lucia.armillotta@anu.edu.au}, Mark R. Krumholz$^{1,2}$ and Yusuke Fujimoto$^{1}$\\
$^{1}$Research School of Astronomy and Astrophysics - The Australian National University, Canberra, ACT, 2611, Australia\\
$^{2}$Centre of Excellence for Astronomy in Three Dimensions (ASTRO-3D), Australia}
\date{Accepted XXX. Received YYY; in original form ZZZ}

\pubyear{2018}

\begin{document}
\label{firstpage}
\pagerange{\pageref{firstpage}--\pageref{lastpage}}
\maketitle

\begin{abstract}
Ongoing surveys are in the process of measuring the chemical abundances in large numbers of stars, with the ultimate goal of reconstructing the formation history of the Milky Way using abundances as tracers. However, interpretation of these data requires that we understand the relationship between stellar distributions in chemical and physical space, i.e., how similar in chemical abundance do we expect a pair of stars to be as a function of the distance between their formation sites. We investigate this question by simulating the gravitational collapse of a turbulent molecular cloud extracted from a galaxy-scale simulation, seeded with chemical inhomogeneities with different initial spatial scales. We follow the collapse from galactic scales down to resolutions scales of $\approx 10^{-3}$ pc, and find that, during this process, turbulence mixes the metal patterns, reducing the abundance scatter initially present in the gas by an amount that depends on the initial scale of inhomogeneity of each metal field. However, we find that regardless of the initial spatial structure of the metals at the onset of collapse, the final stellar abundances are highly correlated on distances below a few pc, and nearly uncorrelated on larger distances. Consequently, the star formation process defines a natural size scale of $\sim 1$ pc for chemically-homogenous star clusters, suggesting that any clusters identified as homogenous in chemical space must have formed within $\sim 1$ pc of one another. However, in order to distinguish different star clusters in chemical space, observations across multiple elements will be required, and the elements that are likely to be most efficient at separating distinct clusters in chemical space are those whose correlation length in the ISM is of order tens of pc, comparable to the sizes of individual molecular clouds.
\end{abstract}




\section{Introduction}
\label{Indroduction}

Understanding galaxy formation and evolution has been a key challenge for astrophysics over the last decades, and, despite remarkable progress, a complete picture of galaxy formation is still lacking. Our Milky Way represents a unique laboratory for this kind of study, and, indeed, several studies have been dedicated to reconstructing its assembly history based on the present-day distribution of gas and stars. Stars, in particular, play a crucial role in this reconstruction due to their power as fossil records of the chemical evolution of our Galaxy. Stars form in clusters within molecular clouds, but most of these clusters survive for much less than a Gyr. The majority are unbound immediately upon formation by expulsion of gas  (e.g., \citealt{Kroupa+02, Baumgardt+07, Fall+10, Murray+10, Kruijssen12} -- see \citealt{Krumholz+14} for a review), while those that survive this initial phase are often disrupted over the next few hundred Myr by processes such as dynamical interactions with other molecular clouds or Galactic tidal fields \citep[e.g.][]{Lada&Lada03, Koposov+10, Dalessandro+15}. Chemical abundances represent, therefore, the only imprint that stars store from their formation site during their entire lives. 

One of the goals of galactic archaeology is to unravel the formation and evolution of the Milky Way by using stellar chemical composition to reconstruct the trajectories of individual stars back billions of years to their birth sites. This technique, proposed by \citet{Freeman02}, is called chemical tagging. The key idea is that stars born from the same star cluster are nearly homogeneous in their chemical abundances, and thus we should be able to use chemical abundances measured with sufficient accuracy to identify stars that originated in the same cluster \citep[e.g.][]{Bland-Hawthorn+10a, Bland-Hawthorn+10b, Mitschang+13, Ting+15}. Reconstructing the dynamical history of the Milky Way through the chemical tagging technique is one of the main goal of recent surveys, such as APOGEE \citep{Zasowski+13,Majewski+17}, the \textit{Gaia}-ESO Public Spectroscopic Survey \citep{Gilmore+12} and GALAH \citep{DeSilva+15}, that observe $\sim 10^5-10^6$ stars with moderate spectral resolution. However, two important requirements must be met in order for chemical tagging to be successful: stars within the same cluster must share similar element abundances, while stars born in different clusters must be characterised by sufficiently different chemical compositions that it is possible to separate them in chemical abundance space. Given its crucial importance for galactic archaeology, both observational and theoretical validations of the central assumptions behind chemical tagging are needed. 

From an observational point of view, the first measurements of abundance variation across multiple element groups in open clusters and moving groups \citep[e.g.][]{DeSilva+06,DeSilva+07,Reddy+12,Ting+12} found scatters around $0.05-0.1$ dex, but this was comparable to the measurement uncertainty. More recent works \citep[e.g.][]{Onehag+14,Bovy16,Liu+16,Ness+18} have improved on previous measurements, yielding robust detections of scatters at the level of $0.02-0.03$ dex. These numbers are significantly smaller than the variation of $0.05-0.3$ dex seen in the interstellar medium (ISM) from which stars form \citep[e.g.][]{Sanders+12, Li+13, Berg+15, Arellano+16, Vogt+17}, thus proving that open star clusters are much more chemically homogeneous than the surrounding environment. While the condition of cluster homogeneity seems to be supported by observations, the requirement of significant cluster-to-cluster variations in chemical abundances is more challenging to validate.  Recently, several works have explored whether it is possible to separate stars that originated in different clusters through their chemical signatures  \citep[e.g.][]{Mitschang+13,Mitschang+14,Blanco-Cuaresma+15,Ting+16}. They have found a high degree of overlap in chemical abundance space: single clumps of stars in chemical space correspond to clusters with masses of $\sim 10^7 \, \mo$ in physical space, suggesting an upper limit to the mass of chemically-homogeneous individual clusters. More accurate data coming from on-going surveys are needed to provide a more precise picture.

While these studies are informative, because they target known clusters with ages $\gg 10$ Myr, they are necessarily restricted to studying the regions of highest star formation efficiency, where enough gas was converted into stars to create a structure that survived gas expulsion. Since only a small fraction of stars form at such sites \citep[and references therein]{Krumholz+14}, it is unclear to what extent the results can be generalised. Moreover, at very young ages stars are not easily divisible into well-defined ``clusters'' and a ``field''; instead, they have a fractal spatial distribution with correlations on all scales \citep[and references therein]{Gouliermis18}. As pointed out by \citet{KrumholzTing18}, metals are similarly likely to be correlated on multiple scales. Given this complexity, it is unclear that a cluster in the chemical sense (i.e., a region of near-uniform chemical abundance) can be identified with the physical open clusters we see at much larger stellar ages -- the cluster that survives as a bound structure today was conceivably part of a much larger-chemically homogeneous structure at earlier times, only the very densest parts of which remained bound. This poses a fundamental problem for the interpretation of chemical tagging data: supposing we do identify a set of chemically homogeneous stars that are now dispersed in space. Should we identify these as having been born within 1 pc of one another, 1 kpc, or some intermediate scale?

These uncertainties create an important role for theoretical studies, which can answer questions that the observations, at present, cannot. The first step to doing so is to determine why, under what conditions, and over what size scales clusters should be chemically homogeneous. The first study in this area was carried out by \citet{Murray+90}. Based on analytical arguments, the authors concluded that turbulent diffusion within a molecular cloud should homogenise its composition in roughly a crossing time. Several years later, \citet{Bland-Hawthorn+10a} and \citet{Bland-Hawthorn+10b} pointed out that a full homogeneity within a cluster must be reached before supernovae start to explode, since even a single supernova can produce a significant change in the chemical pattern. This condition is fully satisfied only in clusters with mass lower than $10^5\,\mo$, suggesting that clusters of this size are certainly suitable for chemical tagging. 
Most recently, \citet{Feng&Krumholz14} performed adaptive-mesh three-dimensional simulations of metal mixing during star cluster formation, starting from artificial ``colliding flow" initial conditions whereby star formation began as a result of a collision between two streams of warm neutral gas with differing chemical compositions. They found that turbulent mixing produces a stellar abundance scatter at least five times lower than the initial gas abundance scatter, confirming that stars in the same cluster are much more chemically well-mixed than the ISM where they form. Moreover, the authors showed that chemical homogeneity is achieved even in gravitationally unbound clusters, confirming the possibility to identify dissolved clusters through star chemical signatures. 

While this result was encouraging, the choice of initial condition meant that the simulation yielded only a single, very large, homogeneous cluster with little substructure. Consequently, the findings of this study do not address the requirement that stars born in different clusters have different chemical composition. Nor could these simulations determine the characteristic size scale, if any, that defines a cluster in the chemical sense. Nor could they study how the level of homogeneity in stars might depend on the spatial structure of the metals at larger scales, since in the colliding flow setup there is only a single spatial scale (the initial separation of the two flows). This is of interest as well, since \citet{KrumholzTing18} have shown that, at galactic scales, elements whose astrophysical origin sites are widely-distributed (e.g., AGB stars) are expected be correlated on smaller scales than elements whose origin sites are rare (e.g., neutron star mergers).

In this paper, we follow up the work of \citet{Feng&Krumholz14}, improving both initial conditions and resolution so that we can study the chemical distribution in newborn stars, and its relationship to their spatial distribution, in a more realistic context. We start from initial conditions for the gas density distribution taken self-consistently from a simulation of an entire galaxy, and then zoom in on a chosen region to simulate the three-dimensional collapse of a giant molecular cloud (GMC) up to resolution of $10^{-3}$~pc, including self-gravity and star formation. The goal of this work is to study how turbulent gravitational collapse influences decay and mixing of chemical fluctuations of varying initial size scales, and how chemical abundances of newly born stars correlate within a star-forming cloud that gives birth to multiple stellar sub-clusters.

This paper is organised as follows. In \autoref{Simulation}, we introduce the initial conditions of our simulation and we briefly describe the main features of the code. In \autoref{Results}, we present an analysis of our results. Finally, in \autoref{Conclusions}, we summarise our work and discuss possible implications for the chemical tagging technique.

\section{Methods}
\label{Simulation}

\subsection{Numerical methods}
\label{Methods}

We perform the simulation described below using the parallel adaptive mesh refinement (AMR) code \textsc{orion} \citep{Li+12}. \textsc{orion} uses a conservative second order Godunov scheme to solve the equations of the ideal magnetohydrodynamics \citep{Mignone+12} coupled to a multi-grid method to solve the Poisson equation for self-gravitating gas \citep{Truelove+98,Klein+99}. In the present simulation we do not include magnetic fields, because the initial conditions from which we begin (see below) do not include them. Radiative heating and cooling are included using the analytic approximation to heating and cooling rates in the neutral ISM given by \citet{Koyama&Inutsuka02}. Although the gas should undergo a transition from atomic to molecular composition during its collapse, we do not include this chemical change in our simulations, nor its effect on the cooling rate. Since the flows we produce are highly supersonic in any event, the $\approx 50\%$ error we make in the cold gas sound speed by omitting the transition from C$^+$ to CO cooling is not dynamically important. We also do not make use of \textsc{orion}'s radiative transfer solver \citep{Krumholz+07,Rosen+17}, since our focus here is not on the origin of the initial mass function or on stellar feedback.

\subsection{Initial Conditions}
\label{InitialCondition}

\subsubsection{First extraction}
\label{FirstExtraction}

To produce a simulation of chemical mixing suitable for the study of stellar spatial distributions over a range of scales, we require realistic initial conditions. We therefore extract the initial conditions of our simulation from the galactic-scale simulation performed by \citet{Fujimoto+16}. The authors of this work simulated the evolution of a M83-like barred spiral galaxy, in order to investigate the effects of star formation and thermal energy feedback on the properties of dense GMCs. The GMCs were defined as coherent structures contained within contours at a threshold particle density of 100~cm$^{-3}$. Based on mass-radius scaling relations, they were classified in Type A, clouds with properties similar to those of observed GMCs, Type B, very massive clouds that have experienced merging, and Type C, clouds with very low density and short lifetimes. 

For our simulation, we select an isolated Type A cloud born in a spiral region at $\sim 5$~kpc from the galactic center. The mass of the cloud, $M_\mathrm{c}$, is $8\, \times \, 10^5\,\mo$, while the average radius, $R_\mathrm{c}$, is 14~pc. The 1D velocity dispersion, $\sigma_\mathrm{1D}$, is $8.1\,\kms$, significantly larger than the sound speed, $1.8 \kms$, so the flows within the cloud are highly supersonic.
The selected cloud, as is the case for most of the isolated clouds in \citet{Fujimoto+16}'s simulation, shows a predominantly two-dimensional geometry. Most of the gas lies on a plane parallel to the galactic plane, with a thickness lower than 10~pc. This particular shape is due to the effect of galactic shear \citep[e.g.][]{Tasker&Tan09,Fujimoto+14}. 

\begin{figure*}
\includegraphics[width=\textwidth]{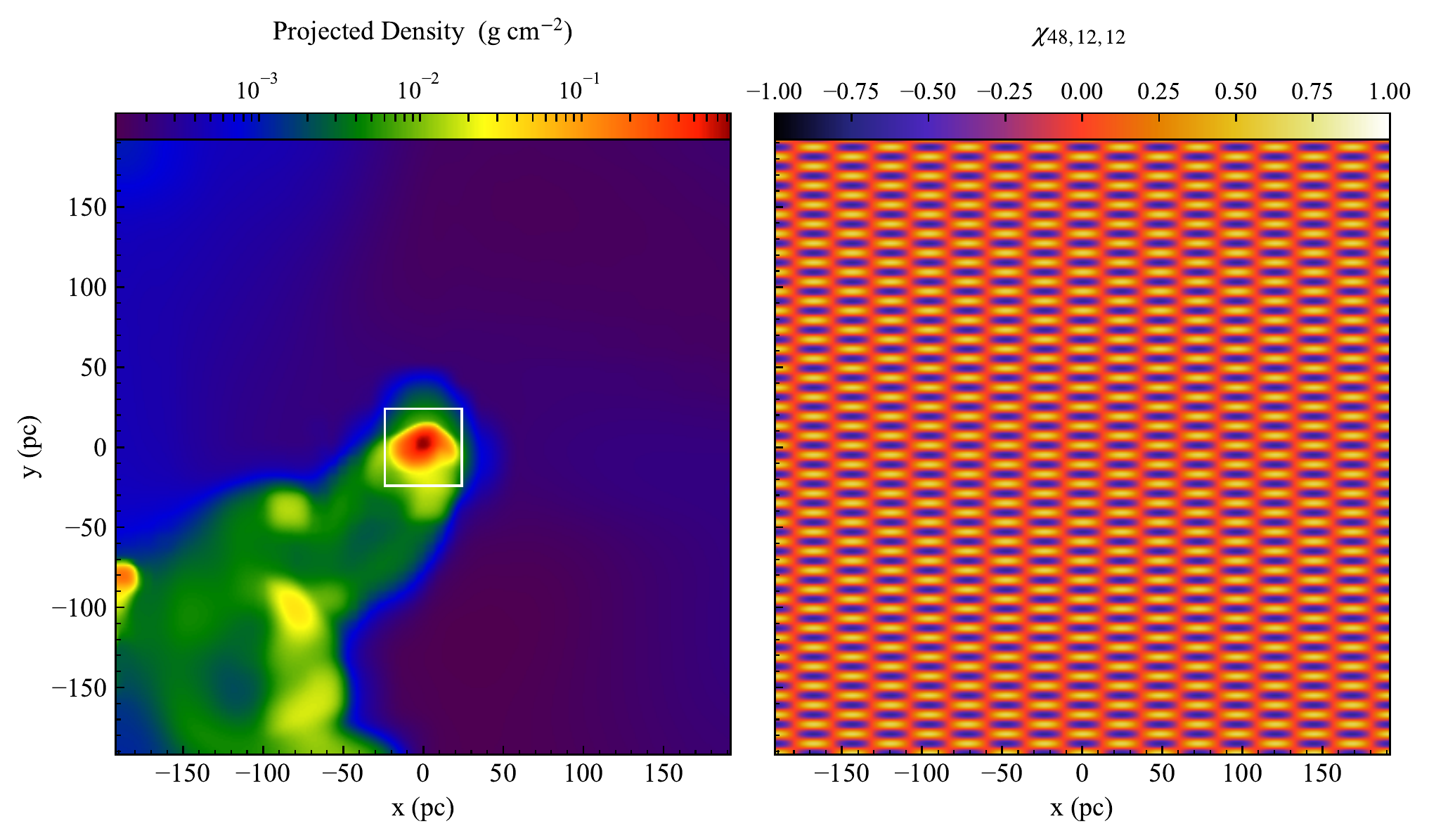}
\caption{Initial conditions of our simulation. \textit{Left panel}: projection of the gas density distribution along the $z$-axis. \textit{Right panel}: distribution on grid of the metal tracer $\chi_{48,12,12}$. The slice is taken at $z\,=\,0$.}
\label{IniCond}
\end{figure*}

In the simulation by \citet{Fujimoto+16}, every cell with mass larger than $1000\,\mo$ was refined of a factor 2 up a maximum resolution of 1.5~pc.  We extract and interpolate to a single resolution all data within a 3D region containing the cloud. The box size of our simulation is 384~pc across each direction, with root resolution of 0.75~pc, i.e., we begin our simulation from a grid whereby we have interpolated the state extracted from the galaxy simulation onto a grid with cells a factor of 2 smaller than the native resolution. The left panel of \autoref{IniCond} shows the initial conditions for the gas density distribution, while the white square encloses the molecular cloud, whose evolution we study at very high resolution (see below). We apply outflow boundary conditions at the edge of our 384~pc box. However, using different boundary conditions would not affect our result, because the molecular cloud is far enough from the grid boundaries that no flows have time to traverse the distance from the grid boundaries to the central molecular cloud during our run time.

Through our simulation, we follow the gravitational collapse of the GMC until its fragmentation into stars. Our goal is to study how the interplay between turbulence and gravity influences the mixing of chemical inhomogeneities. Turbulence within the molecular cloud is fed both by the energy on larger scales gained from the galactic-scale simulation \citep[stellar feedback, gravitational and thermal instabilities, e.g.][]{Yang&Krumholz12, Goldbaum+16}, and by the gravitational energy released during the collapse \citep[e.g.][]{Federrath+11, Goldbaum+11, Robertson+12, Birnboim+18}. The balance between kinetic and gravitational energy within the cloud can be quantified by the virial ratio, which is defined as
\begin{equation}
\alpha_\mathrm{vir}=5\frac{R_\mathrm{c}\sigma_\mathrm{1D}^2}{GM_\mathrm{c}}\;.
\label{alphavir}
\end{equation}  
The typical value of $\alpha_\mathrm{vir}$ for observed star-forming clouds is $\sim 1-2$ \citep[e.g][]{McKee&Ostriker07, Kauffmann+13}. If we enter the cloud parameters, the cloud we select for resimulation has $\alpha_\mathrm{vir}\sim1.3$, in agreement with the observed values. 

\subsubsection{Relaxation phase}

Since the maximum spatial resolution in \citet{Fujimoto+16}'s simulation is 1.5~pc, the hydrodynamical quantities within the cloud are completely smooth on scales of a few times 1.5~pc. As a consequence, the turbulent power spectrum drops off on scales just below the cloud radius. Thus while the cloud is globally in virial equilibrium, its densest regions might not be stable against gravitational collapse due to lack of effective turbulent motions. If we were to suddenly allow the resolution to increase, while allowing the cloud to collapse under self-gravity starting immediately from the initial state after extraction, turbulence would not be able to hinder self-gravity on small scales, causing a collapse of the entire structure rather than a more realistic gradual cloud fragmentation. Even worse, we would underestimate the amount of turbulent mixing that should occur, because gravity would cause the cloud to collapse into stars before the turbulence on previously-unresolved scales had time to reach statistical equilibrium. In effect, we would be starting the collapse from an artificially smooth, non-turbulent state.

To overcome this problem, we use a relaxation procedure whereby we turn on small-scale gravity slowly, in order to ensure that there is adequate time for the turbulent cascade on a small scales to reach statistical steady state before allowing collapse to proceed. Our procedure is as follows. Starting from the state we extract from the \citet{Fujimoto+16} simulation, we soften gravity on distances smaller than eight times the root resolution of our simulation. In practical terms, we solve the Poisson equation on a grid with a resolution of 6~pc covering the entire box and interpolate the solution on finer grids to get the corresponding values of the gravitational potential. The softened gravity provides global confinement that stops the cloud from dispersing and keeps matter accreting onto it, while at the same time preventing small-scale structures from collapsing. During this phase when gravity is softened, we slowly increase the level of refinement in the central 48~pc region containing the cloud (the white square in the left panel of  \autoref{IniCond}). Each time we double the resolution, we run the simulation for a period of time long enough for turbulence to cascade to smaller scales, defined by the condition that the velocity power spectrum approximately follows the Burgers power spectrum \citep{Burgers}, valid for shock-dominated supersonic turbulence, rather than showing a sharp cut-off at the old resolution limit. We repeat this procedure of increasing the refinement level by a factor 2 until we reach a resolution of 0.09~pc in the 48 pc region shown by the white box in \autoref{IniCond}, which is almost two orders of magnitude smaller than the initial cloud radius. 
This procedure takes $ \approx 2 $ Myr of simulation time after the initial extraction.

\subsubsection{Second extraction and star formation}

After 2 Myr, the turbulence has fully relaxed at a resolution of 0.09 pc, and we proceed with the next phase of the simulation. We extract the central 72~pc region containing the cloud and carry on the simulation in the presence of full self-gravity. Outflow boundary conditions are imposed at the edges of our new box. Again, the choice of the boundary conditions is not important for our purposes, because our simulation runs for a time much shorter than the time required for the grid-boundary gas to reach the cloud. In this second part, we increase the resolution in regions undergoing gravitational collapse, refining by a factor 2 any cells in which the local density exceeds the Jeans density, according to the \citet{Truelove+98} criterion
\begin{equation}
\rho_\mathrm{J} = J^2  \dfrac{\gamma \pi k_\mathrm{B} T}{\mu m_\mathrm{H}G \Delta x^2}\;,
\label{Jeans}
\end{equation}
where $J$ is the Jeans number (set to $J=1/8$ in our simulation),  $T$ is the gas temperature,  $\Delta x$ is the cell size, $\mu=1.27$ is the mean molecular weight, and $\gamma = 5/3$ is the ratio of specific heats. These last two values are taken from \citet{Fujimoto+16}'s simulation. We keep them constant since we are not following the transition from atomic to molecular gas.   
We allow 5 levels of refinement, reaching a maximum resolution of 0.003~pc. If the local density on the finest level exceeds the Jeans density (\autoref{Jeans}) with a Jeans number $J = 1/4$, we introduce a sink particle with initial mass
\begin{equation}
M_\mathrm{s, 0} = [\rho - \rho_\mathrm{J}(J=1/4)] \, \Delta x^3\;,
\label{Jeans}
\end{equation}
where $\rho$ is the cell gas density.
In \textsc{orion}, sink particles evolve according to the prescription of \citet{Krumholz+04}: they interact gravitationally with the surrounding gas and accrete mass, momentum and energy from all cells located within an accretion radius $4 \Delta x$ on the finest AMR level. The accretion rate onto sink particles is estimated by fitting the gas around them to Bondi-Hoyle flow. 

For the purpose of our analysis, we assume that each accreting sink particle represents an individual star. We define the star formation efficiency, $\varepsilon_\mathrm{SF}$, as the fraction of mass cloud converted into stars, $M_\mathrm{*}/M_\mathrm{c}$, where $M_\mathrm{*}$ is the total stellar mass and $M_\mathrm{c}$  is the cloud mass defined in \autoref{FirstExtraction}. We stop the simulation when the star formation efficiency reaches a value of $\sim 10 \%$,  consistent with expected maximum star formation efficiencies in GMCs \citep[e.g.][]{Lada&Lada03, Kruijssen12}. 

\subsection{Metal tracers}
\label{Metal tracers} 

The goal of our work is to study how turbulence mixes chemical abundances during the star formation process. However, the simulation performed by \citet{Fujimoto+16} does not include metal fields. In the absence of a self-consistently generated initial metal distribution, we initialise the metal distribution in our simulation by using a series of basis functions, much as in the approach of \citet{Yang&Krumholz12} and \citet{Petit+15}. 

Within our initial, 384 pc box (after the first extraction), we can represent an arbitrary metal field, $\chi (x,y,z)$, as a linear combination of trigonometric functions, $\chi_{\frac{L}{l},\frac{L}{m},\frac{L}{n}} (x,y,z)$, also known as Fourier modes,
\begin{equation}
\chi(x,y,z) = \sum\limits_{l=0}^{\infty} \sum \limits_{m=0}^{\infty} \sum \limits_{n=0}^{\infty} a_{l,m,n} \, \chi_{\frac{L}{l},\frac{L}{m},\frac{L}{n}} (x,y,z)\;,
\label{metalfield}
\end{equation}
where $a_{l,m,n}$ are the Fourier coefficients, that represent the amplitude of each mode. Each Fourier mode is defined by
\begin{equation}
\chi_{\frac{L}{l},\frac{L}{m},\frac{L}{n}} (x,y,z) = \mathrm{cos} \left(2\pi l \dfrac{x}{L}\right) \,  \mathrm{cos} \left(2\pi m \dfrac{y}{L}\right) \, \mathrm{cos} \left(2\pi n \dfrac{z}{L}\right) \;,
\label{Fouriermodes}
\end{equation}
with $L=192$~pc the size of the half box, and \textit{l}, \textit{m}, \textit{n} the wavenumbers characteristic of each mode. Note that, by definition, the value of each Fourier mode can vary between -1 and 1. We only consider cosine functions because the fields represented by sine functions are identical up to translation of $x \mapsto x \pm L/2l$, $y \mapsto y \pm L/2m$, $z \mapsto z \pm L/2n$.

In our simulation, we do not follow the evolution of an arbitrary metal field, $\chi$, but rather the evolution of the single Fourier modes, $\chi_{\frac{L}{l},\frac{L}{m},\frac{L}{n}}$, into which the arbitrary metal field can be decomposed. At the start of the relaxation phase, we add 10 metal tracers following \autoref{Fouriermodes}, where the respective wavenumbers, \textit{l}, \textit{m}, \textit{n}, and wavelengths, \textit{L/l}, \textit{L/m}, \textit{L/n}, are listed in \autoref{Metals}. As an example, the right panel of \autoref{IniCond} shows the initial distribution of the metal tracer $\chi_{48,12,12}$. We highlight that the maximum wavelength, 48~pc, is almost twice the average radius of the cloud. 

\begin{table}
\begin{center}
\caption{List of metal tracers, $\chi_{\frac{L}{l},\frac{L}{m},\frac{L}{n}}$: \textit{l}, \textit{m}, \textit{n} are the wavenumbers, while $L/l$, $L/m$, $L/n$ are the wavelengths characteristic of each metal tracer (see \autoref{Fouriermodes}).}
\label{Metals}
\begin{tabular}{ccccccc}
\hline
$\chi_{\frac{L}{l},\frac{L}{m},\frac{L}{n}}$& $l$&$m$&$n$&$L/l$& $L/m$& $L/n$\\
&&&& (pc) & (pc) & (pc)\\
\hline
$\chi_{48,12,12}$&$4$&$16$&$16$& $48$&$12$&$12$\\
$\chi_{12,48,12}$& $16$&$4$&$16$&$12$&$48$&$12$\\
$\chi_{12,12,48}$& $16$&$16$&$4$&$12$&$12$&$48$\\
$\chi_{48,48,12}$& $4$&$4$&$16$&$48$&$48$&$12$\\
$\chi_{48,12,48}$& $4$&$16$&$4$&$48$&$12$&$48$\\
$\chi_{12,48,48}$& $16$&$4$&$4$&$12$&$48$&$48$\\
$\chi_{48,48,48}$& $4$&$4$&$4$&$48$&$48$&$48$\\
$\chi_{12,12,12}$& $16$&$16$&$16$&$12$&$12$&$12$\\
$\chi_{6,6,6}$& $32$&$32$&$32$&$6$&$6$&$6$\\
$\chi_{3,3,3}$& $64$&$64$&$64$&$3$&$3$&$3$\\
\hline	      
\end{tabular} 
\end{center}
\end{table}

Each metal tracer is treated as a scalar field passively advected by the fluid motion, in the same way as advection of mass.  Note that the conserved quantity in our simulation is not $\chi_{\frac{L}{l},\frac{L}{m},\frac{L}{n}}$, but the metal tracer mass $\int \rho \chi_{\frac{L}{l},\frac{L}{m},\frac{L}{n}} dV$; we conserve this quantity to machine precision. Once sink particles form, they assume the same value of $\chi_{\frac{L}{l},\frac{L}{m},\frac{L}{n}}$ as the gas cell in which they form, so that the total metal mass in gas cells plus sink particles remains constant.

\begin{figure*}
\includegraphics[width=\textwidth]{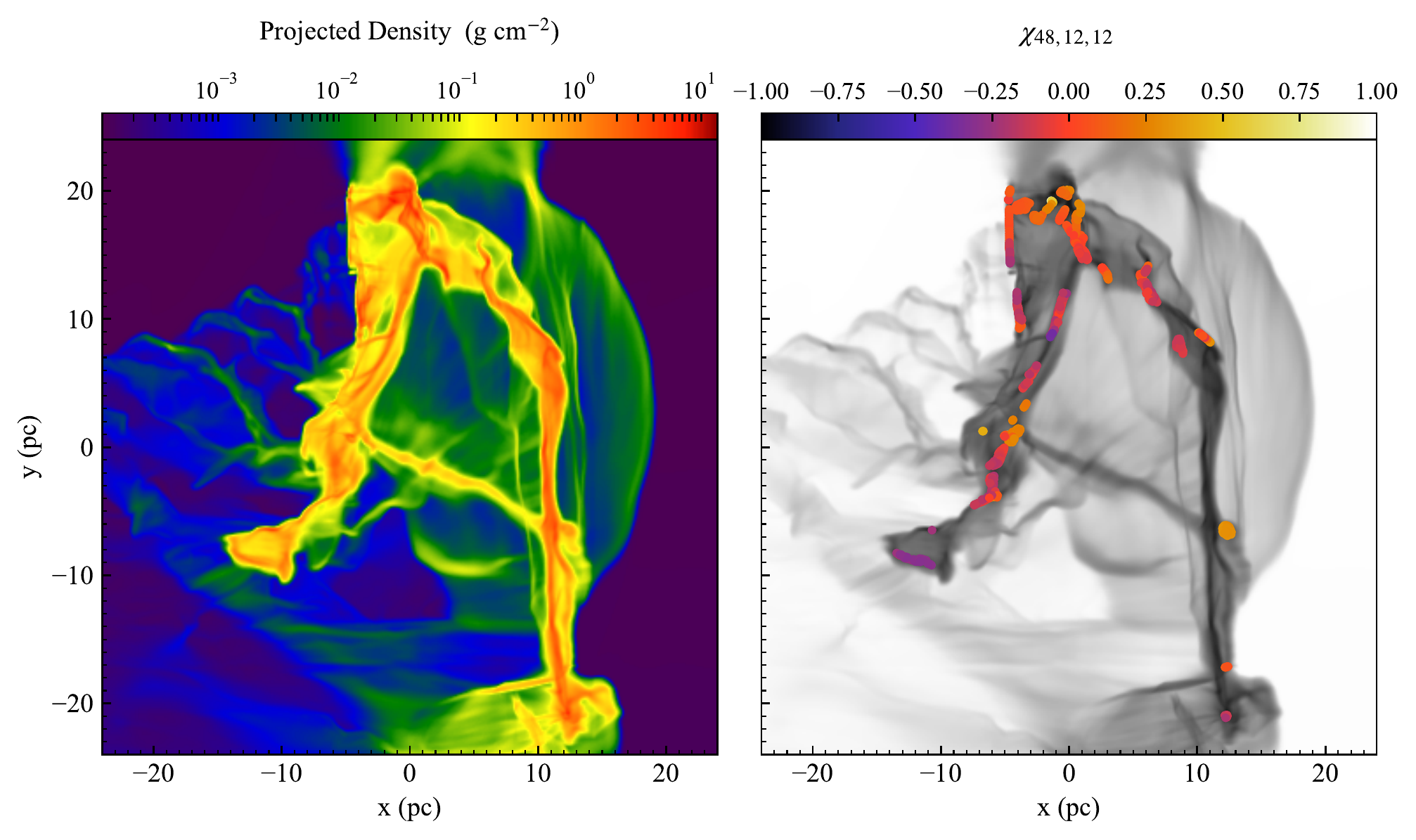}
\caption{ Central 48~pc region at the end of the simulation. \textit{Left panel}: projection of the gas density distribution along the $z$-axis. \textit{Right panel}: star particle distribution (coloured dots) overlapping the projected density distribution (greyscale map). The color of each star indicates the value of the metal tracer $\chi_{48,12,12}$ in that star.}
\label{FinalPlot}
\end{figure*}

In order to track the metal content evolution of each sink particle, we have modified the sink particle algorithm in \textsc{orion} \citep[see also][]{Feng&Krumholz14}. In addition to mass, momentum and energy, sink particles accrete metal mass from each cell located within their accretion radius. At each time step $\Delta t$, the amount of metal mass accreted from a given cell \textit{i} is $\chi_{\frac{L}{l},\frac{L}{m},\frac{L}{n}, i} \,M_{i}$, where $\chi_{\frac{L}{l},\frac{L}{m},\frac{L}{n}, i}$ is the metal tracer value in the cell \textit{i}, while $M_{i}$ is the amount of mass accreted from the cell \textit{i}. Then, the value of each metal tracer incorporated in the sink particle, $\chi^*_{\frac{L}{l},\frac{L}{m},\frac{L}{n}, \mathrm{s}}$, at time $t+\Delta t$ is 
\begin{equation}
\chi^*_{\frac{L}{l},\frac{L}{m},\frac{L}{n}, \mathrm{s}} (t + \Delta t)= \dfrac{\chi^*_{\frac{L}{l},\frac{L}{m},\frac{L}{n}, \mathrm{s}} (t)  M_\mathrm{s} (t) + \sum \limits_{i=1}^{N} \chi_{\frac{L}{l},\frac{L}{m},\frac{L}{n}, i} \,M_{i}}{M_\mathrm{s} (t + \Delta t) }
\label{SinkPartMet}
\end{equation}
where $M_\mathrm{s}$ is the sink particle mass, and $N$ is the number of cells within the accretion area; note that, throughout, we use $\chi^*$ to denote metal fields measured in sink particles, and $\chi$ to denote metal fields measured in the gas.

\section{Results}
\label{Results}

Once we turn on full self-gravity, the simulation takes 8 kyr before stars start to form. We stop the simulation when $\varepsilon_\mathrm{SF}\sim 10\%$, which we reach 27 kyr after the onset of star formation. The final number of stars is $\sim 2000$, with masses ranging between $0.001 \,\mo$ and $500\, \mo$. We emphasise that, because our simulations do not include radiative transfer or radiation feedback, the fragment mass spectrum in our simulations is artificial, and thus the masses of individual particles should not be viewed as reliable \citep{Krumholz+16, Cunningham+18, Guszejnov+18}. Instead of true stars, it is perhaps more accurate to view the sink particles as tracers of the stellar population.

\autoref{FinalPlot} shows the distribution of gas and stars within the cloud at the end of the simulation: the left panel shows the gas density projection along the $z$-axis, while the right panel shows the star distribution overlapping the gas distribution. We can note that the GMC evolves into a network of intersecting filaments. Star formation takes place in the densest parts of these filaments. The average size of each star cluster is a few pc.
 
In the right panel of \autoref{FinalPlot}, the colour of each star indicates the respective abundance of the metal tracer $\chi_{48,12,12}$. We can observe that, while the initial gas abundances vary in a range of values between -1 and 1 (full colour-bar), at the end of the simulation the stellar abundances vary in a much narrower range, roughly half the initial one. This range is further reduced within each star cluster, where the metal field is nearly homogeneous. In fact, gravity leads to the formation of smaller and smaller structures, where turbulence can be quickly dissipated and mixing can easily occur. 

We quantitatively analyse these results in the next sections, while in \autoref{append}, we conduct convergence studies to verify that the results are robust and weakly influenced by numerical resolution.

\subsection{Metal tracer evolution}

As first step, we analyse how metal tracers evolve during the star formation process. For this purpose, we only consider the stellar abundances of each metal tracer, and, at each time $t$, we calculate the Pearson correlation between the current stellar abundances, $ \chi^*_{\frac{L}{l},\frac{L}{m},\frac{L}{n}} (t)$ for stars $s$, and the initial gas abundances in the current star positions $(x_\mathrm{s}, y_\mathrm{s}, z_\mathrm{s})$, $ \chi_{\frac{L}{l},\frac{L}{m},\frac{L}{n}} (x_\mathrm{s}, y_\mathrm{s}, z_\mathrm{s},t = 0)$. Intuitively, this statistic addresses the question: how well do the metal abundances in stars reflect the abundances  that were initially present in the gas at their locations? Is the spatial pattern in the stellar abundances simply a mirror of the initial spatial pattern in the gas abundances (in which case the correlation should be near unity), or does the process of assembling mass to make the stars wash out the gas pattern?

The Pearson correlation between these two discrete data sets is defined by \footnote{Note that, to simplify the notation, we will use $\chi$ to indicate $ \chi_{\frac{L}{l},\frac{L}{m},\frac{L}{n}}$.} 
\begin{equation}
P_\mathrm{\chi} (t) = \dfrac{\mathrm{cov}[\chi^*(t),\chi(0)]}{\sigma_\mathrm{\chi,star }(t)\, \sigma_\mathrm{\chi}(0)}\;,
\label{Pearson}
\end{equation}
where the covariance, $\mathrm{cov}(\chi^*(t),\chi(0))$, is defined by
\begin{equation}
\begin{split}
&\mathrm{cov}[\chi^*(t),\chi(0)] = \\
&{\sum\limits_{s=1}^{N_\mathrm{stars}}  \left \lbrace \left[\chi^* (x_\mathrm{s}, y_\mathrm{s}, z_\mathrm{s}, t)  - \bar{\chi}^*(t)\right]\,\left[\chi (x_\mathrm{s}, y_\mathrm{s}, z_\mathrm{s}, 0)  - \bar{\chi}(0)\right]\right \rbrace}\;,
\end{split}
\label{cov}
\end{equation}
while the scatter, $\sigma_\mathrm{\chi, star}(t)$, is defined by
\begin{equation}
\sigma_\mathrm{\chi, star} (t) = \sqrt{\dfrac{1}{N_\mathrm{star}-1}{\sum\limits_{s=1}^{N_\mathrm{stars}} \left [ \chi^* (x_\mathrm{s}, y_\mathrm{s}, z_\mathrm{s}, t)  - \bar{\chi}^*(t) \right]^2}}\;,
\label{stdev}
\end{equation}
and analogously for the initial gas scatter $\sigma_\mathrm{\chi}(0)$, $N_\mathrm{stars}$ is the number of stars at the time $t$, while $\bar{\chi}^*(t)$ and $\bar{\chi}(0)$ are the mean values of the two distributions.
From \autoref{Pearson}, we can see that the value of $P_\mathrm{\chi}$ can range between -1 and 1, where 1 indicates perfect linear correlation between the two distributions, 0 indicates absence of correlation and -1 indicates perfect anti-correlation. Note, however, that stellar motion could result in a negative correlation even if the underlying metal distribution in both gas and stars were fixed, in which case a decrease of the power of $\chi(0)$ would imply an increase of the power of $\chi_\mathrm{T}(0)$, defined as translation of $\chi(0)$ along the three axes:
\begin{equation}
\chi_\mathrm{T}(x,y,z,t=0) = \chi \left(x + \dfrac{L}{2l}, y+\dfrac{L}{2m} , z +\dfrac{L}{2n},t = 0\right)\;.
\end{equation}
Since we are interested in changes in the spatial pattern of the metals, not simply translations, we calculate the Pearson correlation as
\begin{equation}
P_\mathrm{\chi} (t) = \sqrt{\left(\dfrac{\mathrm{cov}[\chi^*(t),\chi(0)]}{\sigma_\mathrm{\chi,star }(t) \,\sigma_\mathrm{\chi}(0)}\right)^2+\left(\dfrac{\mathrm{cov}[\chi^*(t),\chi_\mathrm{T}(0)]}{\sigma_\mathrm{\chi,star }(t) \,\sigma_\mathrm{\chi_\mathrm{T}}(0)}\right)^2}\;.
\label{Pearson2}
\end{equation}
The newly defined quantity is invariant under translation, and, by definition, it can now range between 0 and 1, where 1 indicates that the initial metal pattern has remained unchanged during the star formation process, while 0 means that turbulence has mixed out the initial inhomogeneities, destroying the original pattern.

\begin{figure}
\includegraphics[width=\columnwidth]{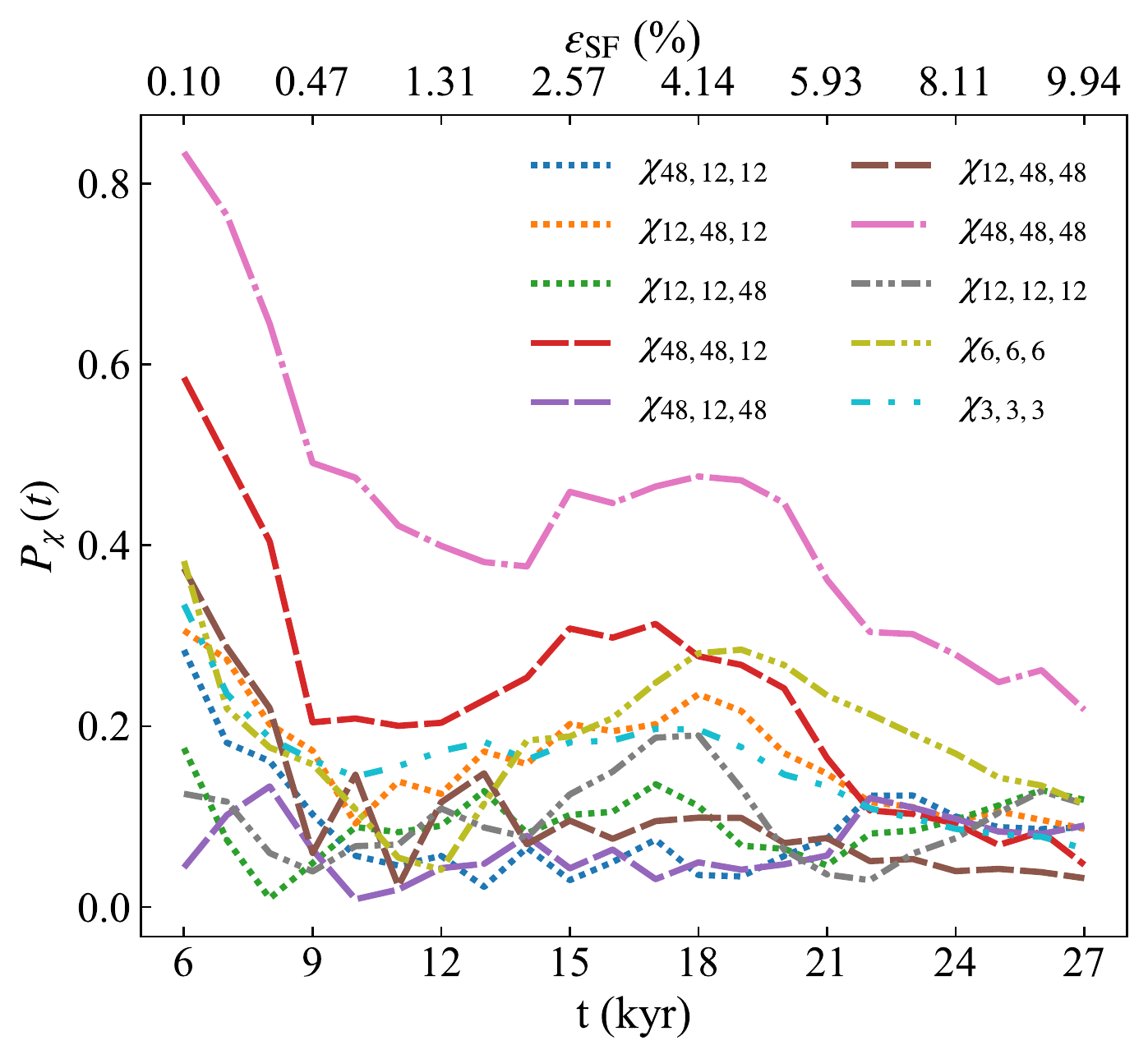}
\caption{Evolution of $P_\mathrm{\chi}$, the correlation between metal abundances in stars and initial metal abundances in the gas at the stars' positions, as a function of time, $t$, and star formation efficiency, $\varepsilon_\mathrm{SF}$, for the 10 different metal tracers. The time $t$ is calculated starting from the onset of star formation.}
\label{TimeEv}
\end{figure}

\autoref{TimeEv} shows the evolution of $P_\mathrm{\chi}$ with time for each mode. Note that we use the same line style to identify modes with the same mean wavelength.  The bottom axis measures the time starting from the onset of star formation, while the top axis indicates the value of the star formation efficiency, $\varepsilon_\mathrm{SF}$, at each time. We begin the plot at 6 kyr, because the amount of cloud mass converted in stars is lower than $0.1\%$ before this time and our sample, composed of fewer than 10 stars, is not statistically robust. 

In \autoref{TimeEv}, we observe that $P_\mathrm{\chi}$ fluctuates between 0 and 0.2 for all the modes except for $\chi_{48,48,48}$ (pink line) and $\chi_{48,48,16}$ (red line), which are modes characterized by fluctuations on large scale. For those two modes, $P_\mathrm{\chi}$ decreases with time, becoming, at the end of the simulation, lower than 0.1 for $\chi_{48,48,12}$ and $\sim 0.2$ for $\chi_{48,48,48}$. We conclude that turbulence efficiently mixes all inhomogeneities on scales comparable with or lower than the cloud size (average diameter in the $xy$-plane $\sim 30$~pc, thickness along the $z$-axis $< 10$~pc, see \autoref{InitialCondition}) before the onset of star formation. Fluctuations on larger scales survive on longer times before being destroyed during the star formation process. 

We conclude this section by noting that, while here we have calculated the correlation between the stars' metal abundances and the initial gas-phase metal abundances at the stars' \textit{current} positions, we obtain qualitatively identical results if we compute the correlation using the stars' \textit{initial} positions (i.e., the locations at which the star particles first formed). Moreover, values of Pearson correlation lower than 0.2 are obtained if we calculate the correlation between the stars' initial metal abundances (i.e., metal abundances when star particles first formed) and the initial gas-phase metal abundances at the stars' \textit{initial} positions. Thus the small Pearson correlation we have measured is a result of turbulent gas flows, not stellar motions.

\begin{figure*}
\includegraphics[width=\textwidth]{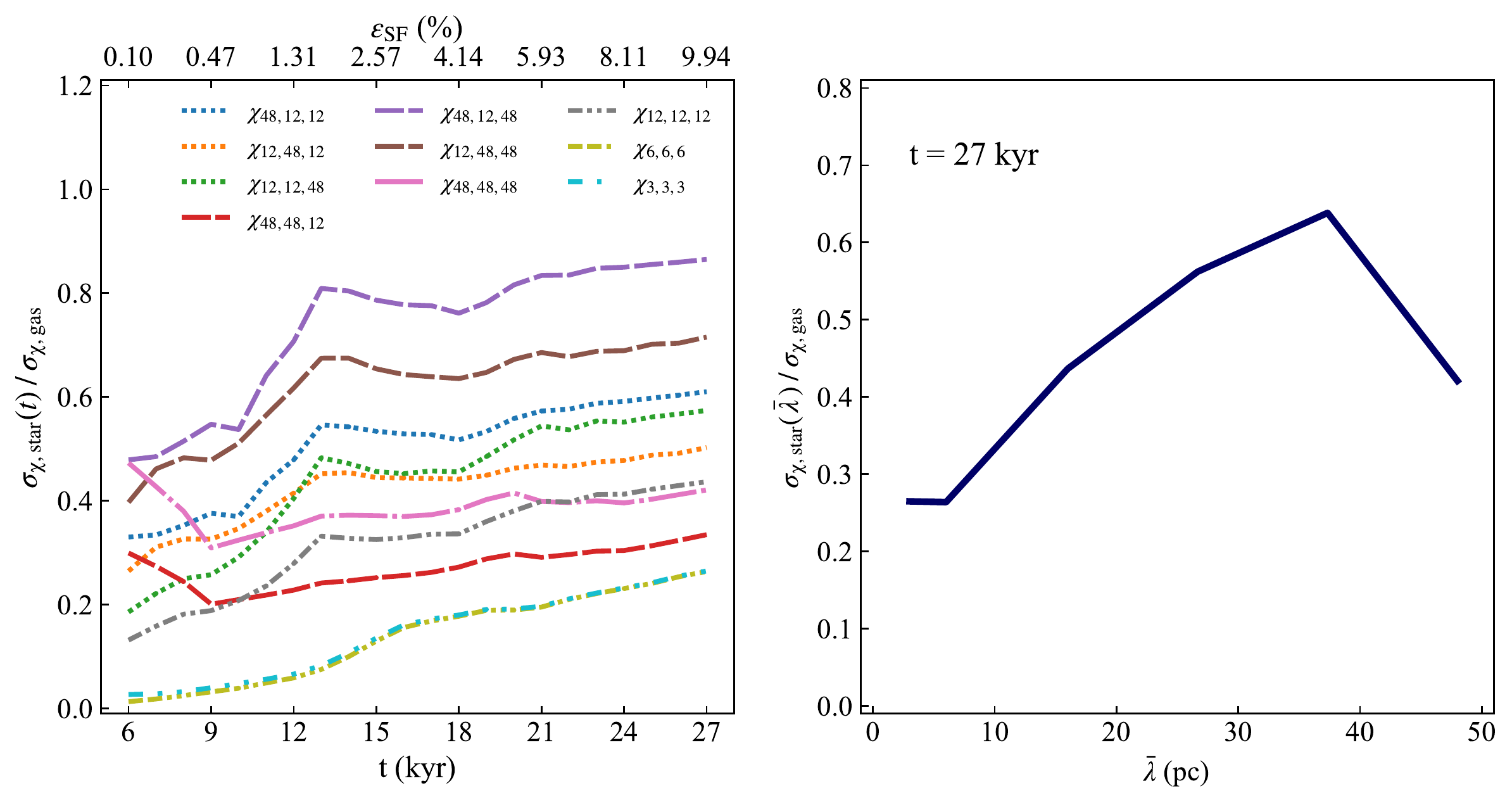}
\caption{Scatter in stellar abundances. \textit{Left panel}: Evolution of the stellar abundance scatter, $\sigma_\mathrm{\chi,star}(t)$, normalized to the initial scatter in gas abundances, $\sigma_\mathrm{\chi,gas} \simeq 0.353$, as a function of time, $t$, and star formation efficiency, $\varepsilon_\mathrm{SF}$, for the 10 different metal tracers. The time $t$ is calculated starting from the onset of star formation. \textit{Right panel}: $\sigma_\mathrm{\chi,star}(t)/\sigma_\mathrm{\chi,gas}$ evaluated at $t=27$~kyr, which is the final time of the simulation, as a function of the mean wavelength, $\bar{\lambda}$, of each mode.}
\label{ScatterPlot}
\end{figure*}

\subsection{Scatter in stellar abundances}
\label{Scatter}

In addition to studying how the abundances in stars correlate with those of gas, it is also of interest to ask how metal abundances vary from star to star. At any time $t$, we can calculate the stellar abundance scatter $\sigma_\mathrm{\chi,star} (t)$ using \autoref{stdev}. 

In order to quantify the decrease of the stellar scatter during the star formation process, we compare it with the scatter in gas abundances evaluated at the beginning of the simulation. 
The initial gas scatter, $\sigma_\mathrm{\chi, gas}$, is the square root of the variance, $\sigma_\mathrm{\chi, gas}^2$, that for the continuous function $\chi$ is defined by
\begin{equation}
\sigma_\mathrm{\chi, gas} ^2= \dfrac{1}{(2L)^3} \int_{-L}^L  \int_{-L}^L  \int_{-L}^L  \mid \chi(\textbf{\textit{x}}, t=0) \mid^2 \, d^3x \;,
\label{variance}
\end{equation}
where the integral has been evaluated across the entire box. Note that $\sigma_\mathrm{\chi, gas}$, is slightly different from the quantity $\sigma_\mathrm{\chi}$ defined in the previous section; both are measures of the abundance variation in gas, but the former is the variance over all points in space, while the latter is the variance between the gas abundance measured at a fixed set of discrete positions specified by the locations of the stars.
For each of our initial Fourier modes, $\sigma_\mathrm{\chi, gas} = \sqrt{1/8} \simeq 0.353$. It is easy to demonstrate that the value of $\sigma_\mathrm{\chi, gas}$ remains unchanged if we evaluate the integral across the central 48~pc region, which is the region where the GMC lies. 

The left panel of \autoref{ScatterPlot} shows the stellar abundance scatter, $\sigma_\mathrm{\chi, star}$, normalized to the initial gas scatter, $\sigma_\mathrm{\chi, gas}$, as a function of time for each mode. As for \autoref{TimeEv}, we begin the plot 6 kyr after the onset of star formation, and the top axis indicates the star formation efficiency as a function of time. The first notable feature of this plot is that the value of  $\sigma_\mathrm{\chi, star}/ \sigma_\mathrm{\chi, gas}$ is always lower than 1. At the end of the simulation, it ranges between 0.2 and 0.85 depending on the metal tracer. Another way to read this result is that, depending on the initial fluctuation scale, turbulent motions within the cloud reduce the abundance scatter initially present in the gas by a factor $1.25-5$.
We can note that, except for $\chi_{48,48,48}$ (pink line) and $\chi_{48,48,12}$ (red line), the scatter quickly increases in the first 13 kyr after the onset of star formation. The reason is that, when star formation begins, there are just a few better-mixed clumps forming stars (pc-size clusters in \autoref{FinalPlot}). As time goes on, star forming clumps located at larger distances from the first few begin to appear, increasing the stellar abundance scatter. However, after 13 kyr, the value of  $\sigma_\mathrm{\chi, star}$ is either constant or very slowly changing with time.

A second notable feature of the plot is that we can easily divide the 10 different metal tracers into groups with the same mean wavelength (different line styles in the left panel of \autoref{ScatterPlot}), momentarily excluding the modes $\chi_{48,48,48}$ and $\chi_{48,48,12}$ from our analysis. We observe that the scatter decreases with decreasing the mean wavelength, confirming that turbulent mixing becomes more efficient on smaller and smaller scales. Differences in the amount of scatter between modes of the same mean wavelength -- for example the $\chi_\mathrm{48,12,12}$, $\chi_\mathrm{12,48,12}$ and $\chi_\mathrm{12,12,48}$ modes (respectively blue, orange and green dotted lines) -- are mostly a reflection of the cloud's close to sheet-like morphology and orientation predominantly in the $xy$-plane, and are smaller than the differences between the scatters for modes of different mean wavelength. The exception to these trends are the modes $\chi_{48,48,48}$ and $\chi_{48,48,12}$. Despite having large wavelengths, they show a very small scatter compared to the other modes. 
The reason for this is simply that these modes produced a small scatter within the cloud from the beginning of the simulation, since their characteristic scales are larger than the cloud size along each spatial direction.

\begin{figure*}
\includegraphics[width=\textwidth]{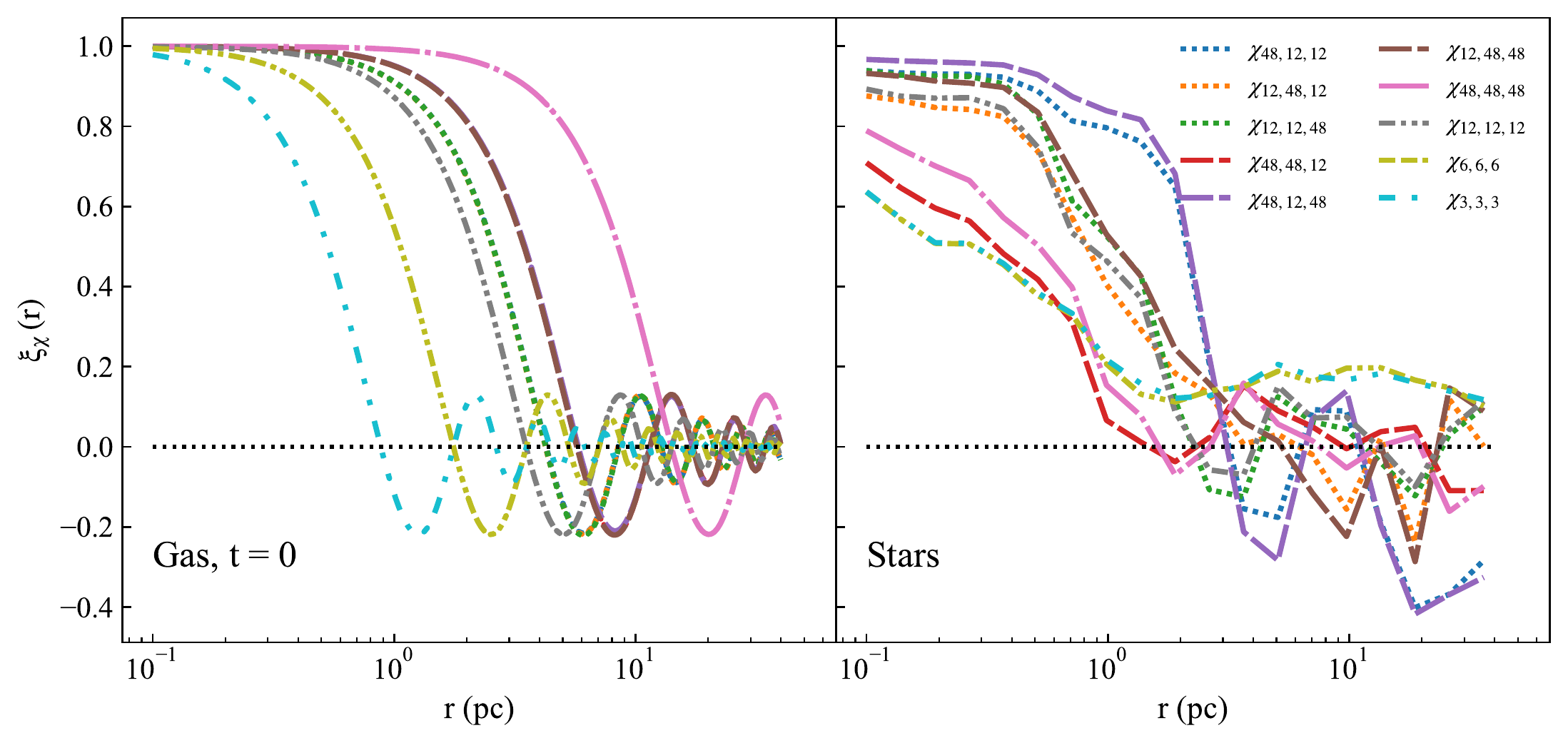}
\caption{Spatial correlation function, $\xi_\chi (r)$, of the gas abundances evaluated at the beginning of the simulation (\textit{left panel}), and of the stellar abundances evaluated the end of the simulation (\textit{right panel}). The correlation function is calculated for the 10 different metal tracers. Note that not all tracer fields are visible in the left panel, because the fields with the same magnitude of spatial scale (e.g., $\chi_{12,48,48}$, $\chi_{48,12,48}$, and $\chi_{48,48,12}$) have identical spatial correlations at the start of the simulation, and thus the lines for them lie on top of one another.}
\label{CorrFunc}
\end{figure*}

The right panel of \autoref{ScatterPlot} allows a faster visualization of all these results. For each mode, we calculate the value $\sigma_\mathrm{\chi,star}$ at the end of the simulation, and we average this value between modes with the same mean wavelength, $\bar\lambda$. We see that the scatter increases with increasing of $\bar\lambda$, reaches a maximum value of 0.64 at $\bar\lambda \sim 40$~pc, and decreases after that. 
We conclude that the abundance scatter is low for small values of $\bar\lambda$ ($ \leqslant 6$~pc) because turbulence efficiently mixes out inhomogeneities on scales significantly smaller than the cloud size and comparable with the size of a single star cluster, and it decreases for large values of $\bar\lambda$ ($\gtrsim 40$~pc)  because fluctuations on these scales are larger than the cloud size, and thus the cloud is nearly homogeneous from the start of its evolution.

\subsection{Spatial correlation of stellar abundances}
\label{correlation}

In addition to the star-star scatter in abundance, we are also interested in the spatial pattern of abundances: how similar are the metal abundances in two stars as a function of the distance between their positions? To answer this question, we examine the spatial autocorrelation function (or simply correlation function) of the stellar abundances for each metal tracer. Calculating the spatial autocorrelation function is equivalent to calculating the Pearson correlation (\autoref{Pearson}) between values of the distribution in different spatial positions, as a function of their relative distance. 

For a continuous field such as our metal tracer distribution in gas, $\chi$, the autocorrelation function is defined by
\begin{equation}
\xi_\mathrm{\chi} (\textbf{\textit{r}}) = \dfrac{1}{\sigma_\mathrm{gas}^2 (2L)^3} \int_{-L}^L  \int_{-L}^L  \int_{-L}^L  [\chi(\textbf{\textit{x}}) \chi(\textbf{\textit{x}}-\textbf{\textit{r}})] \, d^3x \;.
\label{AutocorrC}
\end{equation} 
We are interested in calculating the correlation as a function of the relative distance, \textit{r}, between two stars. Therefore, we calculate $\xi_\mathrm{\chi} (r) = \langle \xi_\mathrm{\chi} (\textbf{\textit{r}}) \rangle$, where $\xi_\mathrm{\chi} (r)$ is function of the magnitude $r = |\textbf{\textit{r}}|$ only, and the angle brackets indicate the averaging over the polar and azimuthal angles. The left panel of \autoref{CorrFunc} shows the spatial correlation function of the gas abundances evaluated at the beginning of the simulation for every metal tracer. The value of $\xi_\mathrm{\chi} (r)$ is equal to 1 for small values of $r$, which indicates perfect correlation. Depending on the metal tracer, $\xi_\mathrm{\chi} (r)$ drops off to negative values at different values of $r$, and then oscillates slowly before converging to zero. Metal tracers with the same wavelength show the same trend (dashed and dotted lines).
Notice that, as expected, the range in separation $r$ over which $\xi_\mathrm{\chi} (r) \approx 1$ for a given metal field is roughly the spatial scale of that field.

To calculate the spatial autocorrelation function for our discrete sample of stellar abundances we proceed using a generalisation of the synthetic catalog methods commonly used to compute two-point correlation functions from galaxy catalogues \citep[e.g.,][]{Coil13}. We first identify all possible pairs of stars within our sample, whose total number is $N_\mathrm{stars} (N_\mathrm{stars} + 1) /2$.  Each pair of stars is characterised by a separation $r_{ij}$, the metal content of the first star $\chi^*_i$, and the metal content of the second star $\chi^*_j$. To estimate the discrete spatial autocorrelation function at separation $r$ in a bin of logarithmic width $\Delta \log r$, we first identify all the pairs $r_{ij}$ such that $|\log(r_{ij}/r)| < \Delta \log r/2$; we use $\Delta \log r = 0.01$ for all the results shown here. We then define the correlation function by the Pearson correlation (\autoref{Pearson}) of the metallicities $\chi^*_i$, $\chi^*_j$ of the identified pairs, i.e.,
\begin{equation}
\xi_{\mathrm{\chi}}(r) = \frac{\sum_{i,j} \left[\left(\chi^*_i - \bar{\chi}^*_i\right)\left(\chi^*_j - \bar{\chi}^*_j\right)\right]}{\sum_{i,j}\left(\chi^*_i -\bar{\chi}^*_i\right) \sum_{i,j}\left(\chi^*_j -\bar{\chi}^*_j\right)},
\label{AutocorrD}
\end{equation}
where $\bar{\chi}^*_i$ and $\bar{\chi}^*_j$ are the means of $\chi^*_i$ and $\chi^*_j$, respectively, for pairs such that $|\log(r_{ij}/r)| < \Delta\log r/2$. We calculate the correlation function for each metal field up to scales $r < 48$~pc, which is the box size of our zoom-in simulation.

The right panel of \autoref{CorrFunc} shows the stellar autocorrelation functions for each metal tracer calculated at the end of the simulation. Qualitatively we see that, not surprisingly, the correlation of metals in stars is high at small spatial separation, and drops to near zero at large spatial separation. Comparing the left and right panels, the most immediately obvious difference between the gaseous and stellar correlations is that the differences in spatial scale of correlation that are obviously present in the initial gas fields have been largely erased in the stellar abundances, so that all stellar fields go from correlated to uncorrelated on roughly the same scale. That is, in the initial gas distribution we have some metal fields that are correlated ($\xi_\chi(r) \approx 1$) out to scales of $\approx 1$ pc (e.g., $\chi_{3,3,3}$, cyan line), while others are correlated at $\gtrsim 10$ pc (e.g., $\chi_{48,48,48}$, pink line). In the stars this range has been compressed, so that all fields go from $\xi_\chi(r) \approx 1$ to $\xi_\chi(r) \approx 0$ over the same narrow range in spatial scale, $r \approx 1-2$ pc.

A second and more subtle feature that one can discern in the right panel of \autoref{CorrFunc} is that the smallest correlations at small separation are found among the fields with the smallest and the largest initial spatial scales, while metal fields that began at intermediate scales show very high correlation at small scales. This is another manifestation of the phenomenon shown in \autoref{ScatterPlot}: initially small-scale modes are uncorrelated because all spatial information has been wiped out by efficient turbulent mixing, while initially large-scale modes show little variance because the star-forming region samples only a small portion of the mode. Intermediate-scale modes produce the most correlation.

\begin{figure}
\includegraphics[width=\columnwidth]{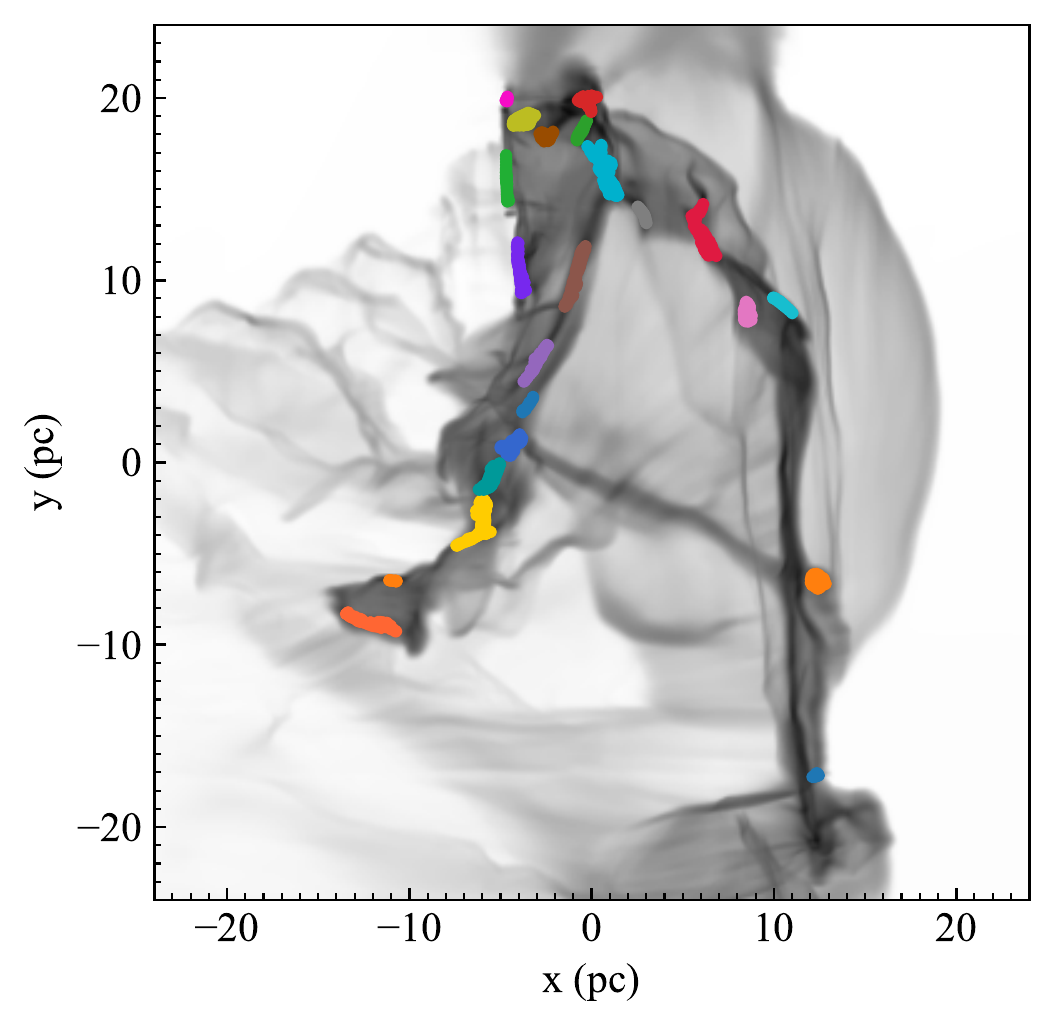}
\caption{Map of the clusters identified by the minimum spanning tree algorithm overlapping the projected gas density distribution (greyscale map, see also \autoref{FinalPlot}). Different colors indicate different star clusters.}
\label{MST}
\end{figure}

\subsection{Statistics of individual star clusters}

The primary finding from \autoref{correlation} is that, independent of the initial spatial scale of a metal field, once stars form their abundances in that metal will be highly correlated on scales below a few pc, and nearly uncorrelated on larger scales. Thus there is a natural size scale of $\sim 1$ pc that defines a chemically-homogeneous star cluster. Motivated by this result, we now seek to investigate the statistics of individual star clusters defined on this scale.

As a first step in our analysis, we use the minimum spanning tree (MST) algorithm to identify individual pc-scale star clusters at the end of our simulation. MST requires that we choose a pruning length to break the stars into sub-clusters, and there are many possible ways to choose this scale \citep[e.g.][]{Schmeja11}. Based on our finding in the previous section that there is a natural size scale of $
\approx 1-2$ pc for chemically homogenous clusters (since the correlation function goes from near unity to near zero sharply at this size scale), we hand-tune our pruning length to identify structures of this characteristic size. After some experimentation, we find that the pruning length that best identifies pc-scale structures is 0.2~pc, so we adopt this value. We set a minimum group size of 10 stars, thus finding 22 groups of stars, with only $4 \%$ of the total amount of stars excluded by this analysis. The map of the identified clusters is displayed in \autoref{MST}.

\begin{figure*}
\includegraphics[width=0.85\textwidth]{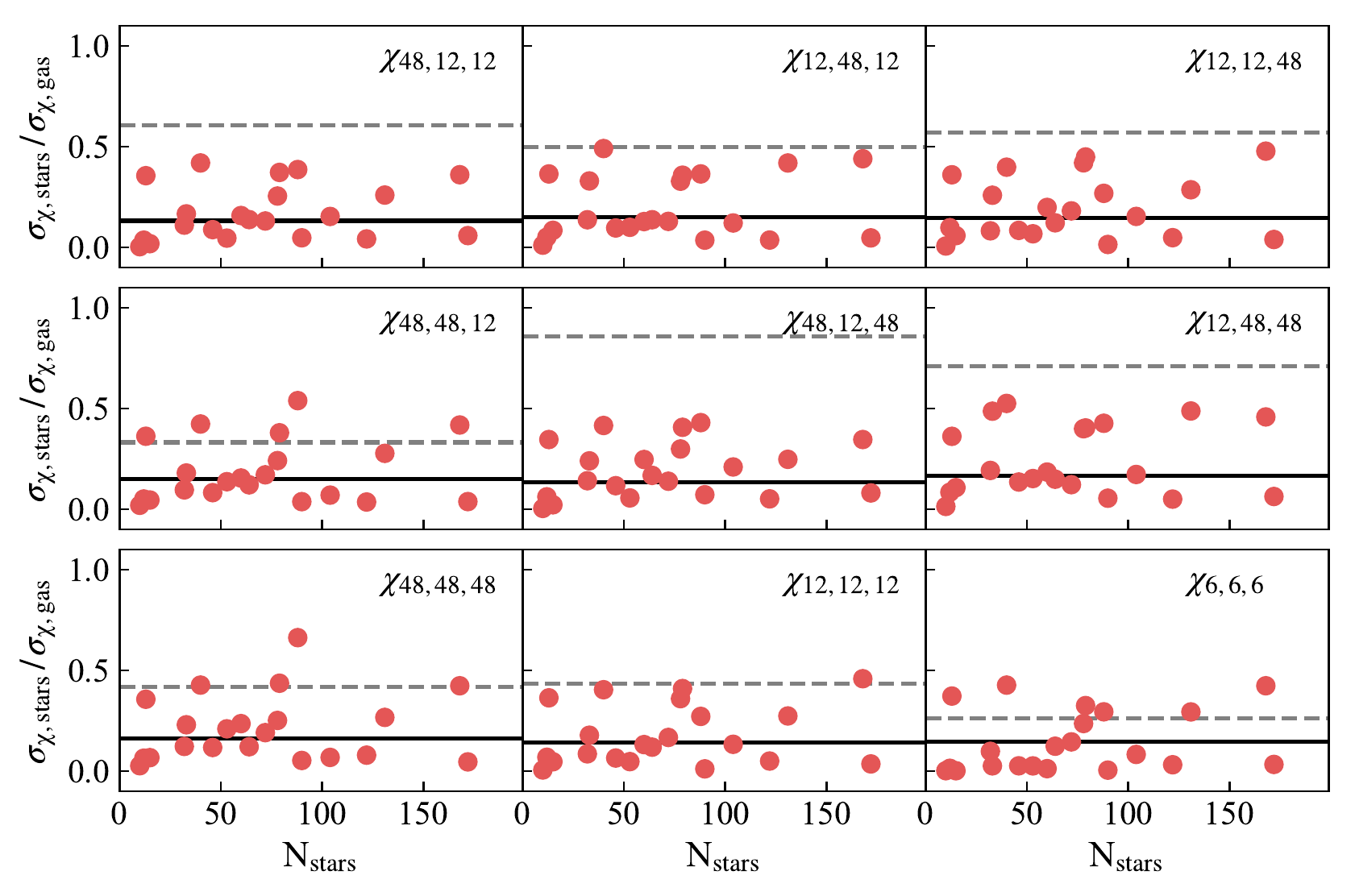}
\caption{Stellar abundance scatters within individual pc-scale star clusters (red dots) as a function of their number of stars, $N_\mathrm{stars}$. The black line indicates the average scatter, while the grey dashed line indicates the scatter between all the stars within the GMC. Different panels correspond to different metal tracers. We do not display the stellar abundance distribution for the mode $\chi_{3,3,3}$ because it is very similar to the one of the mode $\chi_{6,6,6}$.}
\label{ClumpScatter}
\end{figure*}

In \autoref{ClumpScatter}, we show the stellar abundance scatters, $\sigma_\mathrm{\chi,stars}$, within individual clusters (red dots) as a function of the number of stars in the cluster. Different panels correspond to different metal tracers. As in \autoref{ScatterPlot}, the scatter is normalized to the initial scatter in gas abundances,  $\sigma_\mathrm{\chi,gas}$, evaluated across the entire 48~pc cloud region.  
Regardless on the metal tracer and the number of stars within each cluster, the scatter within an individual 1 pc-sized cluster ranges between 0.001 and 0.5~$\sigma_\mathrm{\chi,gas}$. The mean value of these intra-cluster scatters (black solid line) is always lower than the scatter evaluated over all the stars formed in the simulation (grey dashed line), and, as a direct consequence, it is significantly lower than the initial scatter in gas abundances. This is simply a reflection of the fact that the clusters are highly-correlated in metallicity, so each has an internal scatter much smaller than the scatter over all stars. On average, the abundance scatters within single pc-scale clusters are reduced by a factor $\sim 7$ compared to the initial abundance scatter on GMC scales. This result further proves that mixing is very effective in these high-density structures.

\begin{figure*}
\includegraphics[width=0.85\textwidth]{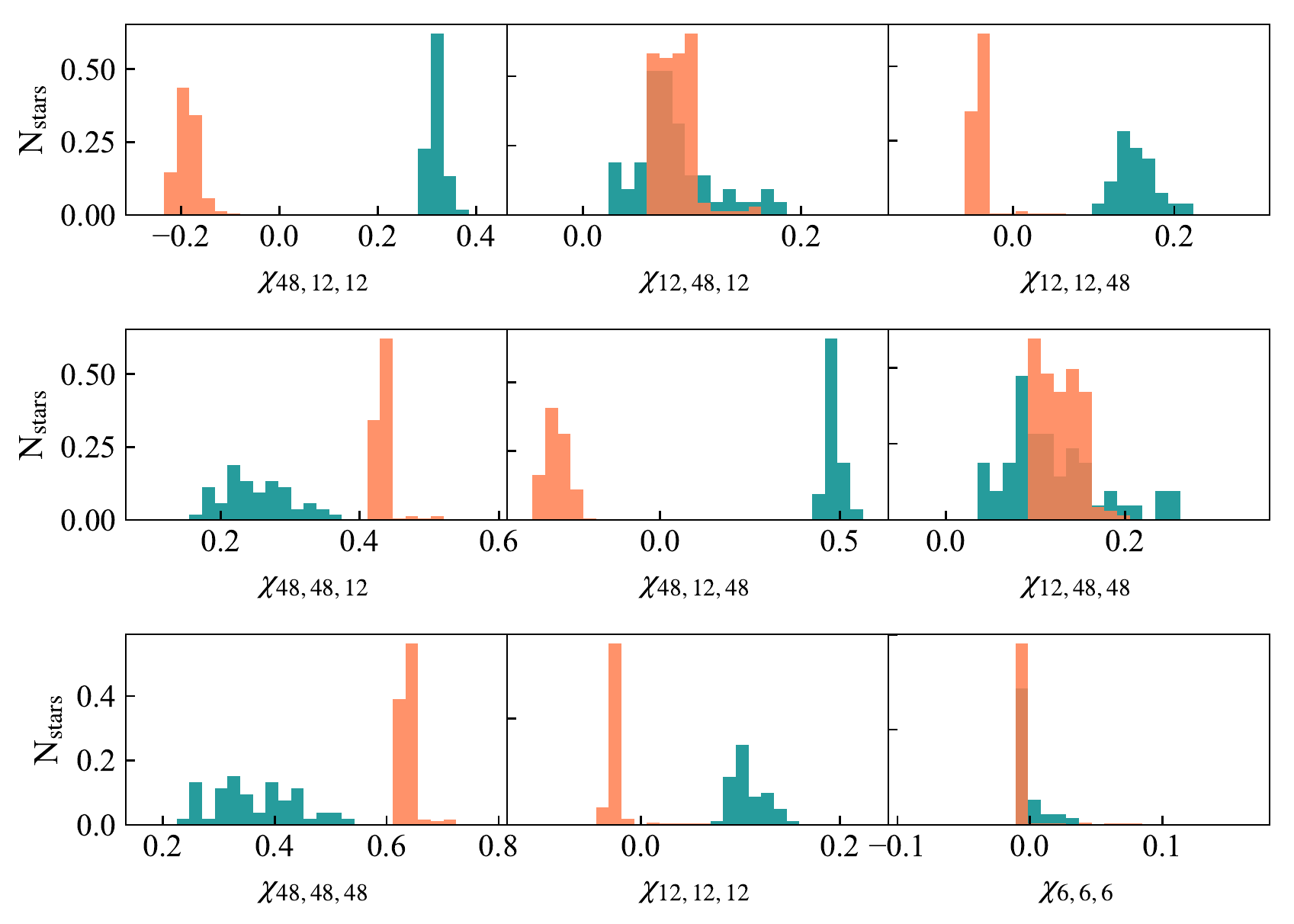}
\caption{Metal tracer distribution for the stars in two neighbouring star clusters, indicated by the green and orange histograms. The histograms are normalised such that the area under them is equal to 1. Different panels correspond to different metal tracers.}
\label{ClumpMetalDistr}
\end{figure*}

\begin{figure}
\includegraphics[width=\columnwidth]{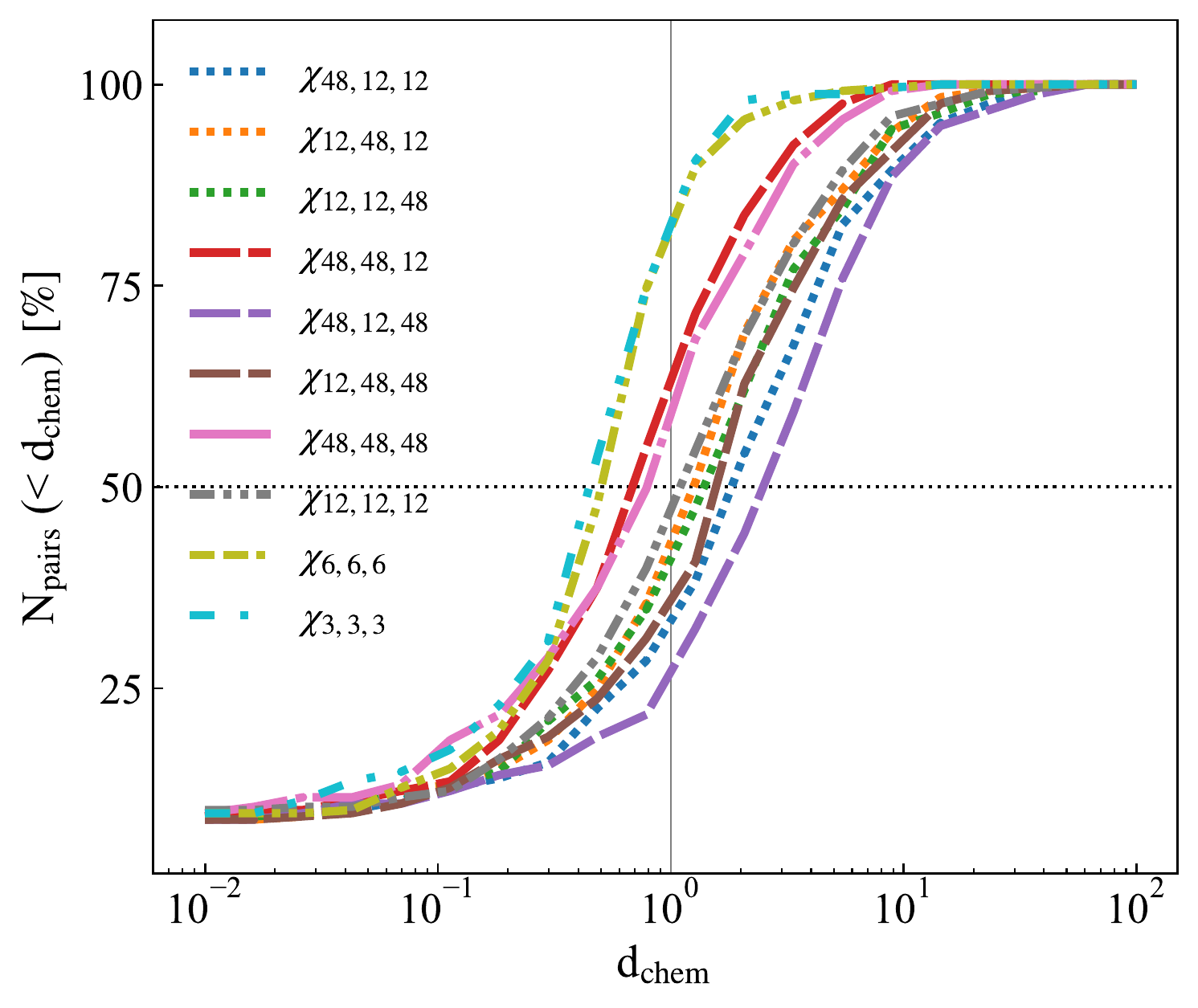}
\caption{Cumulative distribution of chemical distances, $d_\mathrm{chem}$, for all the pairs of star clusters present in our simulation. Different coloured lines distinguish the distributions of the 10 different metal fields. The solid gray vertical line separates the regions of plot where $d_\mathrm{chem}$ is $<$ and $>1$, while the black dotted horizontal line corresponds to half-height of the distribution.}
\label{ChemDist}
\end{figure}

To illustrate the nature of the difference between intra-cluster and inter-cluster scatter, as an example in \autoref{ClumpMetalDistr} we display the distributions of each metal tracer in a different panel for a pair of clusters separated form one another by a few pc. The two differently-coloured histograms indicate the distributions of metal abundances for the stars in the two different clusters. Consistent with our statements above, each distribution shows a very narrow profile compared to the initial metallicity range (between -1 and 1). Depending on the metal tracer, the two distributions can be clearly distinct in chemical space (e.g. $\chi_\mathrm{48,12,12}$ and $\chi_\mathrm{12,12,48}$) or completely overlapping (e.g. $\chi_\mathrm{12,48,12}$). Whether the clusters are chemically distinct or overlapping depends both on the initial level of homogeneity and on the mixing history of a given metal tracer in the region where star clusters form. Therefore, we can conclude that, while each individual pc-scale cluster is characterised by a high level of chemical homogeneity, the uniqueness of their chemical signatures only holds for some tracers. 

The fact that different clusters are chemically-separable in some metal fields but not others suggests that, in order to distinguish different clumps in chemical space, observations of multiple elements will be required. It is therefore of interest to investigate how the degree of separation of two star clusters in chemical space varies depending on the metal tracer selected for analysis, since this can inform the choice of elements used for chemical tagging.  Thus, for each metal tracer, $\chi$, we define the chemical distance, $d_{\mathrm{chem}, ij}$, between two star clusters, \textit{i} and \textit{j}, as follows
\begin{equation}
d_{\mathrm{chem}, ij} =  \dfrac{\left|\bar{\chi}^*_{i} - \bar{\chi}^*_{j} \right|}{\sqrt{\sigma_{\mathrm{\chi,star}, i}^2+\sigma_{\mathrm{\chi,star}, j}^2}}\;.
\label{dchem}
\end{equation} 
where $\bar{\chi}^*_{i}$ and $\bar{\chi}^*_{j}$ are the mean stellar metallicities within each cluster, while $\sigma_{\mathrm{\chi,star}, i}$ and $\sigma_{\mathrm{\chi,star}, j}$ are the scatters. Values of $d_{\mathrm{chem}, ij} \gtrsim 1$ indicate that the two distributions are clearly distinct in chemical space, since the difference between their mean values is larger than the quadrature sum of the scatters. \autoref{ChemDist} shows the cumulative distribution of chemical distances evaluated over all the pairs of star clusters present in our simulation. We note that, for all the metal tracers, the distributions range across almost four orders of magnitude, confirming the requirement of multiple elements to separate two star clusters in chemical space. However, the fraction of star-cluster pairs characterized by chemical distances above 1 significantly varies with the metal tracer: it is $< 50 \%$ for the fields with the smallest and the largest initial fluctuation scales,  while it is $>50 \%$ for the fields with intermediate scales. 
This result reflects the same trend of \autoref{ScatterPlot}, where the abundance scatter between intra-cloud stars is small only if the initial metal-field scale is either significantly smaller or larger than the cloud size.

We also investigate the presence of a possible correlation between chemical and physical distances, but we do not find any correlation between the two in our simulation. The mean abundances of clusters are essentially random with respect to one another as soon as their separation is large enough that they can be identified as discrete clusters. This is not surprising, since we have already found that the star-to-star correlation is essentially zero for separations above $\sim 1$ pc.

\section{Discussion and Conclusions}
\label{Conclusions}

In this work, we perform an adaptive-mesh high-resolution simulation to follow the turbulent mixing of chemical inhomogeneities during the collapse of a star-forming GMC. The initial conditions for the gas density distribution are extracted from a galactic-scale simulation, while the metal fields are initialized by using a series of Fourier functions. Each metal tracer is thus characterized by three wavelengths along the three spatial directions, and evolved as a passive scalar field. We choose values of wavelengths that range from physical scales larger than the cloud size to scales comparable with the size of the smallest and densest structures within the cloud. This approach permits us to investigate not only the distribution of metal abundances in stars, but also its dependence from the different spatial structures that different metals exhibit on galactic scales. 

The final structure of the cloud is the result of the interplay between turbulence and gravity. Gas compression induced by turbulence-generated shocks breaks the cloud into gravitationally unstable filaments, where star formation can occur. We use statistical tool to analyse the distribution in star abundances of the different metal tracers. 
The major findings of our work are as follows:
\begin{enumerate}
\item The turbulent collapse efficiently mixes out the initial metal patterns within the GMC, making the scatters in star abundances smaller than the scatters initially present in gas abundances (see \autoref{ScatterPlot}). Overall, we observe that the scatter in star abundances reaches its maximum value for metal fields with initial fluctuation scales comparable with the cloud size, while it is very small for metal fields with initial fluctuation scales both much smaller and larger than the cloud size.
\item Regardless of the metal field, the stellar abundances correlate on physical scales below a few pc (see right panel of \autoref{CorrFunc} and \autoref{ClumpScatter}). Thus, while the chemical patterns present on cloud-size scales are completely random and uncorrelated, individual pc-scale star clusters display a very high level of chemical homogeneity. Thus the star formation process appears to pick out a size scale of $\approx 1$ pc as the natural one that defines a chemically-homogeneous star cluster. Moreover, while the abundance scatters among all stars formed in the same $\sim 30$ pc-scale molecular cloud are related to the the initial metal field structure, the scatters within individual pc-sized clusters do not.
\item Different pc-scale star clusters may be distinguished in chemical space by using abundance distributions of multiple metals; use of multiple tracers is necessary, because for any single tracer a given pair of clusters may or may not have overlapping abundance distributions (see \autoref{ClumpMetalDistr}). Analysing the metal distributions across all the star cluster pairs present in our simulation, we find that, on average, metals with initial fluctuation scales comparable with or slightly smaller than the GMC size are the most effective at separating pc-sized homogeneous clusters in chemical space, since these metals are characterised by the largest scatters across the cloud (see \autoref{ChemDist}).
\end{enumerate}

There results may have important implications for chemical tagging studies. They confirm the leading role of turbulence in homogenising the chemical composition of a GMC during the star formation process. Moreover, they help to shed light on the possibility of mapping stars between chemical and physical space, i.e., the possibility of determining how closely together in space two stars formed based on the level of similarity or difference in their chemical composition. We remind that the reader that the chemical tagging technique can be successful only if one is able to infer something about the physical distance between the birth sites of two stars starting from their observed separation in chemical space (see \autoref{Indroduction}).

Our results suggest that the characteristic physical scale of individual star clusters in chemical space is a few pc. However, in order to distinguish different physical star clusters in chemical space, observations of multiple elements with different galactic-scale spatial distributions are required. The ISM from which stars form exhibits chemical inhomogeneities across a wide range of physical scales. \citet{KrumholzTing18} show that the scale of spatial correlation of different elements may vary based on their origin sites, e.g. Type II and I supernovae, AGB or neutron stars. Among them, elements produced by AGB stars are expected to correlate on length scales lower than 100~pc, while elements produced by supernovae or mergers should be significantly correlated even on length scales of $\approx 0.5-1$~kpc. According to our results, we might speculate that AGB-generated elements are particularly suitable for chemical tagging studies, since their fluctuation scales are comparable with the size of star-forming GMCs.

In conclusion, chemical tagging needs to operate in a multi-dimensional chemical space in order to be effective. Identifying homogeneous star clusters in chemical space as pc-scale physical structures might be feasible thanks to ongoing surveys (e.g. GALAH, \textit{Gaia}-ESO), that will measure the stellar abundance distribution in our Galaxy across a wide range of chemical elements.

\subsection{Future prospects}

We end by discussing possible next steps in this work, with the ultimate goal of producing results that might be directly compared to the observations of stellar chemical abundances coming from the ongoing star surveys. The first step in this respect is to perform simulations with realistic initial conditions for the gas chemical state generated self-consistently in galactic-scale simulations. These simulations will allow us to investigate the distribution of a wide range of chemical elements - from $\alpha$ to iron-peak and $s$ and $r$-process elements - within star clusters.

Moreover, an accurate characterisation of the star formation process requires the presence of physical ingredients that we have neglected in the present work. First of all, in future simulations we plan to include stellar feedback and radiative transfer. As discussed in \autoref{Results}, thermal support enhanced by stellar radiation feedback is a key mechanism in setting the cloud fragmentation scale \citep[e.g.][]{Krumholz+16}. This step will allow us to trace a more realistic cloud fragmentation history into individual stars. 
Secondly, we plan to include the presence of magnetic fields in our simulations. Recently, \citet{Birnboim+18} have shown that magnetic fields reduces the dissipation rate of kinetic energy during cloud gravitational collapse and, therefore, their presence may play a non-negligible role in turbulence mixing inside the cloud. Inclusion of these effects will have the additional advantage of rendering the star formation process self-regulated, so that there will be no need to stop the simulation according on external criteria, as we have done in the present work.

\section*{Acknowledgements}
The authors acknowledge support from the Australian Government through the Australian Research Council's Discovery Projects funding scheme (project DP160100695). The simulation was performed on the Raijin clusters at the National Computational Infrastructure (NCI), which is supported by the Australian Government.




\bibliographystyle{mnras}
\bibliography{biblio}



\appendix
\section{Convergence study}\label{appendix}
\label{append}

To verify the robustness of our results, we conduce convergence tests at different resolutions. In particular, we focus on the evolution of the scatter in gas abundances (see \autoref{Scatter}). The process of star formation lasts only 35 kyr, however, the entire simulation is much longer, 2.035 Myr. Resolving the mixing of each metal tracer before the onset of star formation is important to have accurate results of stellar abundances.

\begin{figure}
\includegraphics[width=\columnwidth]{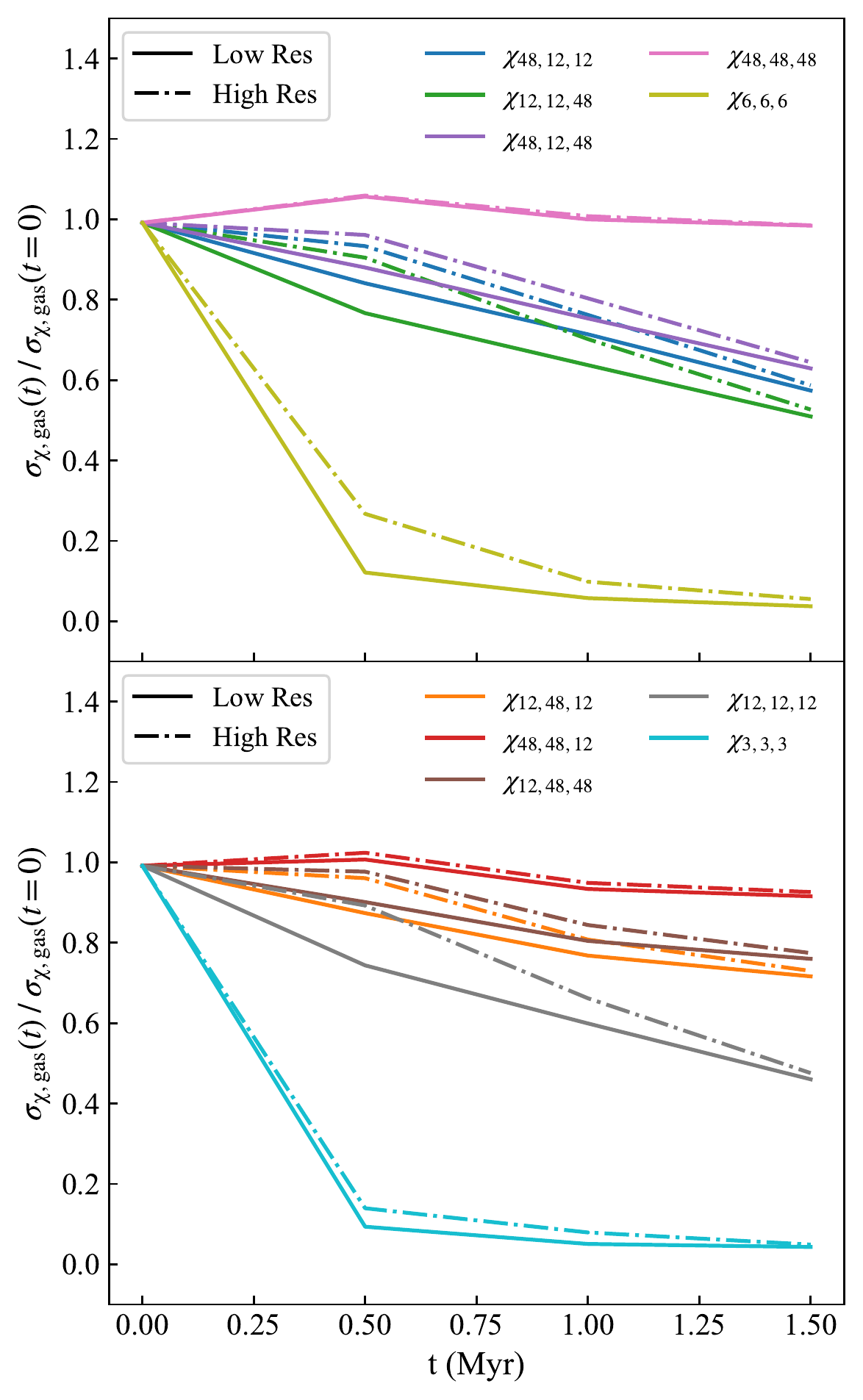}
\caption{Evolution of the gas abundance scatter, $\sigma_\mathrm{\chi,gas}(t)/\sigma_\mathrm{\chi,gas}(0)$ for the 10 different metal fields. The solid and dash-dotted lines indicate respectively the evolution for the run at low (initial resolution 0.75~pc) and high resolution (initial resolution 0.375~pc).}
\label{ResTest}
\end{figure}

In \autoref{InitialCondition}, we explain that in the first part of our simulation the resolution of the central 48~pc region slowly increases from 0.75~pc to 0.09~pc. Here, we perform two simulations, the first one with a coarse resolution of 0.75~pc (as in our reference simulation), the second one with a coarse resolution of 0.375~pc. We run the simulations for 1.5 Myr, increasing the resolution in the central region by a factor 2 every 0.5 Myr. 

\autoref{ResTest} shows the evolution of the gas abundance scatter for each metal field as a function of time. The scatter has been calculated only in the central 48~pc region, which is the region enclosing the star-forming cloud. We observe that the modes $\chi_{48,48,48}$ (pink lines) and $\chi_{48,48,12}$ (red lines) display similar trends at the two different resolutions. For all the other modes, the two trends diverge during the first 0.5 Myr. The scatter at low resolution (solid lines) is always lower than the one at high resolution (dash-dotted lines), because the effect of numerical diffusion of smoothing any fluctuation is stronger at low resolution. However, the two trends show an increasing convergence after 0.5 Myr, as we increase the resolution. At 1.5 Myr, the scatter at the two different resolutions is almost the same for all the modes. This might be due to a better-solved turbulence at high resolution, that makes the mixing process more efficient.


\bsp	
\label{lastpage}
\end{document}